\documentclass[a4paper,usenatbib,useAMS]{mnras}
\usepackage{epsfig,amssymb,amsmath,rotating,lineno,graphicx}
\usepackage[T1]{fontenc}
\usepackage{cleveref}
\usepackage{ae,aecompl}
\usepackage{xcolor}
\usepackage{gensymb}
\usepackage{makecell}
\usepackage{lastpage}
\newcommand\asat{{\it AstroSat}}
\newcommand\asca{{\it ASCA}}

\newcommand\nustar{{\it NuSTAR}}
\newcommand\xmm{{\it XMM-Newton}}
\newcommand\swift{{\it SWIFT}}

\title[constraining corona in NGC 4593]{Understanding coronal geometry in NGC 4593 using Fourier frequency-resolved covariance and time-lag spectral analysis } 

\author[Shree Suman et al.]
{Shree Suman$^{1}$, Mayukh Pahari$^{1}$, Gulab Dewangan$^{2}$, Ian M McHardy$^{3}$\\
$^{1}$ Department of Physics, Indian Institute of Technology, Hyderabad, Kandi, Sangareddy 502285, India \\
$^{2}$ Inter-University Centre for Astronomy and Astrophysics (IUCAA), Pune, 411007, India \\
$^{3}$ University of Southampton, University Road, Southampton SO17 1BJ, UK} 

\begin{document}

\pagerange{\pageref{firstpage}--\pageref{lastpage}} \pubyear{2025}

\maketitle

\label{firstpage}

\begin{abstract}
Understanding disc-corona geometry through X-ray reverberation variability studies in Seyfert galaxies is crucial, yet our knowledge mostly relies on flux-averaged mean spectral analysis. In this study, we investigate the origin of the large X-ray variability of the Seyfert 1 galaxy NGC 4593 using two \xmm{} observations, which are at least 65 ksec long and have a 0.3-10 keV X-ray flux difference by a factor of $\sim$2.5. We extracted mean spectra, Fourier-frequency resolved covariance, and time-lag spectra and performed modelling of all spectra in a self-consistent manner. From the best-fit covariance spectra, we have shown that energy-dependent covariance during low flux shows dominances of direct powerlaw continuum over reflection continuum at all Fourier frequencies (2.1 $-$ 390 $\times$ 10$^{-5}$ Hz). However, during high flux, the variabilities are dominated by the reflection components most of the time. Our results are further supported by the Fourier frequency-dependent time-lag (between soft: 0.3-1 keV and hard: 1-5 keV bands) spectral modeling during high and low fluxes. A significant change is observed in the X-ray reverberation delay timescale from 483 $\pm$ 135 sec (during high flux) to $<$96 sec (during low flux), indicating a change in coronal size at least by a factor of $\sim$2 (from $<$3.3 R$_g$ to $>$7.2 R$_g$) during low to high flux transitions.
\end{abstract}


\section{Introduction}
The X-ray continuum observed in Seyfert galaxies is thought to be dominated by the power-law continuum, which is inverse Compton scattering of soft photons of the optically thick and geometrically thin disc by the hot electron plasma, known as corona. Although the inner accretion disc is assumed to be geometrically thin, optically thick, and neutral \citep{sh73}. Despite such a simple, widely accepted description of inner accretion geometry, the origin of many spectral features is yet to be understood satisfactorily.
For example, several attempts have been made to understand what drives the origin soft excess \citep{b93,GD04,pa20}, the smooth rise in spectra below 2 keV, although competitive hypotheses exist. High-resolution spectral analysis of such soft excess indicates two possible origins: (1) soft excess is a part of the reflection continuum, which shows relativistically smeared, enhanced X-ray emission below 2 keV, \citep{ro05,cr06,wa13} (2) thermal Comptonization of UV photons in the inner part of the disc, known as warm Comptonization \citep{ma98,gi04,do12,xu21}. 
Therefore, mean spectral analysis, even with a sophisticated and realistic physical model, may not be sufficient to understand the significant flux variability. 

Many studies have relied on conventional estimators like fractional root mean square (rms) variability, $F_{var}$ and excess variance, $\sigma^2_{xs}$ \citep{UT09}, which may yield inconsistent results due to their sensitivity to noise and underlying processes in the light curves. The application of frequency-resolved covariance spectral modelling \citep{SM03} offers a novel approach to analysing the relationship between different wavelengths and their corresponding variabilities. This technique can help to understand the origins of variable components and reflection feature components in Seyfert galaxies. However, application of such a technique requires observations of bright and variable X-ray sources with sufficiently long exposure, which NGC 4593 fulfills.  
Being a variable Seyfert 1 galaxy, NGC 4593 has been the subject of extensive X-ray studies, revealing a soft excess in its spectrum as observed by {\it EXOSAT} \citep{po88}. Subsequent observations with {\it BeppoSAX} confirmed a broad absorption dip below 1 keV, which is likely due to a warm absorber along the line of sight \citep{ge98,bl05}. \asca{} data further indicated the presence of a mildly broadened Iron line near 6.4 keV, along with evidence of warm absorption \citep{na97,br07}. Using \xmm{} observation, \citet{br07} noted a time lag of $\sim$230 sec between the soft band (0.5-0.84 keV) and hard continuum (1.5-10 keV), which they interpreted as the finite scattering time within the Comptonizing medium coupling disc and corona.
In particular, significant variability in flux has been noted; for instance, between two \asca{} observations spaced 3.5 years apart, the flux in the 2--10 keV band increased by approximately 25\%. The variability spectra of NGC 4593 suggest that emission components, such as power-law and reflection components, can be correlated. 

To understand the relation between direct and reflected continuum in this source, several attempts have been made in optical/UV/X-ray correlated variabilities studies. Using thermal reverberation from a standard Novikov-Thorne accretion disc, illuminated by an X-ray point-like source, \citep{ka23,ka21} shows that, except for an excess in the U-band, the broadband lag spectra can be well described by such geometric configurations. Using X-ray, FUV and NUV observations from \asat{}, \citet{ku23} observed far- and near UV photons are delayed with respect to X-ray photons by $\sim$38 ks and $\sim$44 ks, favouring a delayed reprocessed emission from the accretion disc. Using an intensive \swift{} campaign, \citet{mc18} observed that the delay of U-band photons with respect to X-rays is exceedingly higher than that expected from the light travel distance between disc and corona; however, filtering of long, slow variabilities from UV and X-ray lightcurves, restored the expected X-ray/UV reverberation time scales. Using \swift{} and \xmm{} joint campaign, with a 140 ksec exposure, \citet{be23} noted that UV photons are delayed to X-ray photons by $\sim$23 ks when long-term variabilities (longer than 5 days) are removed from both lightcurves. using \swift{} and HST lag measurements \citep{ca18}, they noted that lag time scales are consistent with disc-reprocessing with lamp-post geometric configurations of disc-corona. However, \citet{zh21} noted that a geometry more complex than the lamp-post corona may be required to model photon-energy-dependent time lag in a few Seyfert galaxies, including NGC 4593.
Therefore, an independent mean spectral analysis tool is required to understand variabilities in different flux scales in NGC 4593 while maintaining consistency with the results from long-standing mean spectral analysis.   

Among a few popular approaches, the rms spectrum technique \citep{va97,SM03} is particularly useful for studying these correlated variations as it isolates variable components from the overall spectrum. If the geometry of the system allows for the observed power law to produce reprocessed emission through X-ray heating of the accretion disk, both components should manifest together in the variability spectra. However, modelling of RMS spectra has not been explored.

To delve deeper into the origins of variability in NGC 4593, our study employs Fourier frequency-resolved techniques like rms, covariance and time-lag between hard, direct continuum and soft reprocessed continuum to analyse their behaviour within the variability spectra. By separating high-frequency and low-frequency components, we can explore how these variations correlate with underlying physical processes. 
The primary objective of this study is to investigate the origins of X-ray variability in NGC 4593 on different time scales by employing frequency-resolved covariance spectra and time lag spectral modelling on two distinct observational datasets, which are long, variable and have count rate differences by a factor of 2-3. We have calculated photon energy-dependent RMS and covariance spectra as a function of Fourier frequencies, which primarily indicate the presence of a strongly variable spectral component and other non-variable components. We fit them using the model component similar to that used in mean spectral modelling, primarily focusing on a comparative study between direct and reflected continuum. The covariance spectral analysis has been supported by frequency-dependent time lag spectral modelling during low and high fluxes. Lag spectral modelling provides a quantitative idea of the changes in inner disc-corona geometry during the transition from high to low flux. By analysing how X-ray emissions fluctuate on different timescales as well as different energies, this research aims to provide deeper insights into the processes driving variability in this AGN. Section \ref{sec2} provides a data reduction method, while mean, covariance and time-lag spectral analysis are provided in Section \ref{sec3}. Results are discussed, and their implications are provided in Section \ref{sec4}, while conclusions are provided in Section \ref{sec5}.

\section{Observation and Data Reduction}\label{sec2}
NGC 4593 was observed multiple times by \xmm{}/EPIC-pn camera. we have used the two longest observations taken during May 10-11, 2002 (OBD ID: 0059830101; WITH an exposure of $\sim$87 ksec), and July 19-21, 2016 (OBS ID: 0784740101; with an exposure of $\sim$142 ksec). We have chosen these two observations for two reasons: (1) Long exposure observations are required to compute covariance and lag spectra at very low Fourier frequencies. (2) They have significantly different X-ray flux and variabilities in 0.3-10 keV energy ranges (see Figure \ref{fig1}). Hence, they are most suitable for current analysis.   
The data are reduced using the \xmm{} {\textsc{Science Analysis Software (SAS) version xmmsas202304121735-21.0.0} and the Current Calibration File (CCF) {\textsc XMM-CCF-REL-412} on 11 October 2024. For each observation, we run the {\textsc epproc} pipeline. We applied the standard selection criteria with the {\tt FLAG $==$ 0} and {\tt PATTERN $\leq$ 4} condition. Good Time Intervals (GTIs) are calculated by excluding the period for which 10.0-12.0 keV background rate is above 0.4 counts per second for {\textsc EPIC-pn} with the {\tt PATTERN $==$ 0}. Such GTIs have also been used to extract filtered datasets. Source and background lightcurves for both observations have been extracted in 0.3-10 keV from a filtered dataset with the time bin sizes of 128s, 256s, 512s and 1024s, respectively. In all analyses, wherever applicable, background-subtracted lightcurves are used. The left panel of Figure \ref{fig1} shows 0.3-10 keV background-subtracted lightcurves of both observations using a 256 sec bin size.  To ensure a robust comparison between high and low flux source properties, we adopt a conservative approach in selecting lightcurves: (1) the chosen high- and low-flux intervals are well separated in flux/count rate, (2) absolutely no overlap exists between the two intervals, and (3) both have equal durations, allowing a direct comparison across Fourier frequency ranges. As observed in the left panel of Figure \ref{fig1}, the first 25 ksec, where overlap occurs, is discarded from both datasets, leaving 65 ksec of the 2002 lightcurve, matched with a 65 ksec segment from 2016. We also make sure that no data gaps exist in either observation during the selected 65 ksec so that timing studies are not affected. Selected lightcurves are shown in the right panel of Figure \ref{fig1}. This uniform selection enables consistent comparison of covariance, mean spectra, and time lags throughout the analysis. 
To test the selection effect, however, we have analysed the low flux data during the last 100 ksec of the observation in 2016. By including longer segments of the light curve, we were able to probe lower Fourier frequencies for low flux only, and the average count rate increased from 7.4 counts/sec to 8.6 counts/sec. The statistics improved marginally in mean and lag spectral analysis. However, no changes that could significantly impact our results and conclusions have been observed.
Respective GTIs have been created, and filtered datasets have been extracted. Lightcurves on the top right panel of Figure \ref{fig1} show selected data which have been further used in the rest of the analysis. Since 2002 and 2016 observations maintain high and low flux throughout 65ksec, we have used the terms `high flux' and `low flux' in all figures, tables and text to indicate results belonging to 2002 and 2016 observations, respectively.
Using a 65 ks filtered dataset, the source and background spectra were extracted for two circular regions with a diameter of 800 in the physical coordinate. Both regions are chosen so that there is no overlap between them, and they are sufficiently separated so that the background region has negligible contribution from source emission. The auxiliary response files (ARFs) and redistribution matrix files (RMFs) for each spectrum were generated with the \textsc{rmfgen} and \textsc{arfgen} tools. Spectral channels were grouped so that each bin contained at least 30 counts. All spectra are fitted using \textsc{XSpec V. 12.14.1}.

\begin{figure*}
\centering
\includegraphics[scale=0.3,angle=-90]{fig1a.ps}
\includegraphics[scale=0.3,angle=-90]{fig1b.ps}
\caption{Both panel shows \xmm{}/EPIC-pn 0.3-10.0 keV light curve of NGC 4395 during 2002 (left panel) and 2016 (right panel) observations with 128s bin size. Red circles in both panels show 65 ksec segment selections of lightcurves, which have well-separated count rates during high and low flux observations and were used for further analysis in this work. }
\label{fig1}
\end{figure*}

\section{Analysis and Results}\label{sec3}
To explore different features of the mean energy spectra of high and low flux observations and compare them with the rms and covariance spectra at different Fourier frequencies, we provide below details of extractions, computations, analysis and modelling for clarity on steps for the reproduction of results.  
\subsection{Mean Spectral Analysis and results}
Mean spectral modelling was performed separately using spectra extracted from 65ks data during high and low flux separately (as shown in the top right panel of Figure \ref{fig1}).  When the 0.3–10 keV EPIC-pn spectrum of NGC 4593 is fitted with a power-law model (\textsc{zpowerlw} in \textsc{XSpec}), a pronounced soft excess ($<$1 keV) is observed in the residual along with the influence of a warm absorber (top left panel of Figure \ref{fig2}), indicated by a notch-like feature in the 0.7–1.1 keV range. All residual plots used in this work are normalised by statistical errors, assuming Poisson data.
Since the fundamental nature of such soft excesses remains a topic of debate, it is crucial to analyse this feature in detail. Following \citet{mc18,br07}, we have used two partially-ionized warm absorbers described by \textsc{zxipcf$_1$} and \textsc{zxipcf$_2$}. \textsc{zxipcf} model is defined by the column density, the ionisation parameter, and the partial covering fraction.
\begin{figure*}
\centering
\includegraphics[scale=0.3,angle=-90]{high-flux-plaw.ps}
\includegraphics[scale=0.3,angle=-90]{fe-line-both.ps}
\includegraphics[scale=0.3,angle=-90]{spec-new-high.ps}
\includegraphics[scale=0.3,angle=-90]{spec-new-low.ps}
\caption{Top left panel shows the average energy spectra of NGC 4593 fitted only with a powerlaw, along with its residual. A soft excess component, an absorption and a Fe line complex can be seen in the residual. To elaborate on the nature of the Iron line complex, we have zoomed the fitting residual in 4-8 keV (top right panel) when low and high flux spectra are fitted only with a power-law. The bottom left panel shows best-fit spectra during high flux, along with model components and residuals, while the bottom right panel shows the same during the low flux observations.}
\label{fig2}
 \end{figure*}
In addition, the residual shows two distinct emission features: the fluorescent K$\alpha$ line of neutral iron at 6.4 keV and the Ly$\alpha$ recombination line of Hydrogen-like iron at 6.97 keV (top left panel of Figure \ref{fig2}). For clarity, both features, along with the emission and absorption complexes (near 7.6 keV), are shown in the top right panel of Figure \ref{fig2} during high and low flux. We may note that the Fe K$\alpha$ line during low flux is stronger than that during high flux. The neutral iron line probably originates from fluorescence in response to hard X-ray irradiation from a hot corona in the inner accretion disc, occurring in the outer layers of the accretion disc, optically thick broad emission line clouds, or the ``molecular torus" \citep{ma91}. In contrast, the ionised iron line may be produced by irradiation of an ionised disc or through radiative recombination in highly ionised, outflowing material above the disc. In order to model reflection features, including cold, narrow Fe emission lines, we have used the \textsc{xillver} reflection model \citep{ga10,ga13}. \textsc{xillver} self-consistently includes the Fe emission line complex as well as the soft excess if we assume reflection as the origin of the soft excess. The model also takes into account the angular dependence of the emitted radiation and includes a primary power law with an exponential cut-off of up to 1 MeV. We have tied the photon powerlaw index of \textsc{xillver} with that of \textsc{zpowerlw}, and cutoff energy and ionisation parameters are kept free to vary. 
An absorption-like feature is observed at nearly $\sim$6 keV. When such a feature is taken care of by the Gaussian-shaped absorption model, \textsc{gabs}, $\chi^2$/dof has improved from 195/150 to 171/147 (nearly 2.8$\sigma$). Such an absorption feature has been observed before \citep{LA13} and can be an instrumental/calibration feature (e.g., feature due to Fe$^{55}$ calibration source) since no astrophysical absorption lines are reported at this energy.
With the final model (M1): \textsc{zxipcf$_1$*zxipcf$_2$*(zpowerlw*gabs+xillver)} the low flux spectra provide an acceptable fit with $\chi^2$/dof = 171/147 (1.16). 
However, the same model, combination of continuum powerlaw, cold reflection, and double absorption, cannot provide an acceptable fit for the high-flux spectrum with $\chi^2$ / dof = 266/147 (1.81). An excess in the residual has been observed below 2 keV. 

\subsubsection{On the reflection origin of soft excess}
Two major approaches have been extensively used in the literature to model the origin of the excess residual below 2 keV or soft excess: (1) warm Comptonization, where low-energy disc photons are upscattered from the hot, optically thick corona, and (2) high-energy photons from the corona are reprocessed from the disc and provide a reflection continuum with an excess at low X-ray energies. 
We have tested both hypotheses. We have added \textsc{nthcomp} to the M1 model as a proxy for warm comptonization to describe the soft excess and obtained an fit with $\chi^2$/dof = 197/145 (1.36). 
In an alternate approach, we have replaced the \textsc{nthcomp} with the \textsc{xillver} model to M1 to describe the soft excess as a result of reflection. Replacing \textsc{nthcomp} with \textsc{xillver} also provides an acceptable fit with $\chi^2$/dof = 201/144 (1.39). Therefore, adding an unblurred reflection model with the cold reflection plus powerlaw continuum improves the fit statistics ($\chi^2$/dof  change from 266/147 to 201/144).
To describe soft excess in NGC 4593, we prefer reflection over a warm Comptonization scenario based on the following observations. 
We have assessed the nature of Fe K$\alpha$ and K$\beta$ lines during low and high fluxes by fitting 3-9 keV spectra with \textsc{zpowerlw}, and the residuals are shown in the top right panel of Figure \ref{fig2}. The Fe K$\alpha$ line appears to be stronger during low flux compared to high flux. For a more quantitative assessment of such a difference, we have used two \textsc{zgauss} models to describe both emission lines. 
During high flux, the Fe K$\alpha$ line energy and the equivalent width in the rest frame are $6.41^{+0.01}_{-0.02}$ keV and $116^{+24}_{-18}$ eV while the same line during the low flux has the equivalent width of $246^{+33}_{-25}$ eV respectively. We have calculated flux due to the \textsc{zgauss} model describing 6.41 keV Fe K$\alpha$ line using \textsc{cflux} model. During high flux observation, Fe K$\alpha$ line flux is found to be 4.9$^{+0.5}_{-0.5}$ $\times$ 10$^{-13}$ ergs s$^{-1}$ cm$^{-2}$ while the same during low flux is found to be 6.7$^{+1.1}_{-1.2}$ $\times$ 10$^{-13}$ ergs s$^{-1}$ cm$^{-2}$. Errors on fluxes and equivalent widths are quoted with 2$\sigma$ significance. Therefore, the Fe K$\alpha$ line flux increases during the high-to-low flux transition with at least 2$\sigma$ significance. 
Similarly, the rest-frame energy of the second line is $6.98^{+0.02}_{-0.04}$ keV during high flux, while the same is $6.91^{+0.03}_{-0.06}$ keV during the low flux. The equivalent width of the second line is $55^{+13}_{-14}$ eV, while the same during low flux is $91^{+26}_{-21}$ eV, respectively. 

The Fe line observed in $\sim$6.9-7 keV range is consistent with Fe XXVI emission due to ionised Iron. Since the origin of such a line resulting from ionised reflection, their changes in equivalent widths indicate a possible outer disc origin rather than distant neutral material like NLR. 
Therefore, we prefer to use the blurred reflection model to describe the soft excess observed in high flux. Such a choice is also supported later by the fact that we have detected reverberation lag between the hard continuum and the soft excess.
We have used \textsc{xillver$_{sx}$} along with \textsc{kdblur3} as the additional component to model the soft excess. {kdblur3} is a relatively simple blurring function that uses a double broken power-law emissivity profile \citep{wi11}. The blurred reflection model with respect to unblurred reflection, improves the fit moderately. The $\chi^2$/dof improves from 201/144 to 177/141.
The best-fit model for high flux is found to be (M2): \textsc{zxipcf$_1$*zxipcf$_2$*(zpowerlw*gabs+xillver+xillver*kdblur3)}, which provides an acceptable fit with $\chi^2$/dof = 177/141 (1.25).
The photon index of \textsc{xillver$_{sx}$} is tied to that of the \textsc{zpowerlw} model. As the \xmm{}+\nustar{} joint spectral modelling prefers a high cut-off energy ranging between 170-800 keV \citep{ur16}, at different fluxes, the cut-off energy is kept fixed at 300 keV in our analysis.
The inclination angle is kept fixed at 35$\degree$ \citep{re04}. A significant change in the disc ionisation parameter is observed from low flux ($\log \xi$ = $0.36^{+0.24}_{-0.11}$) to high flux ($\log \xi$ = $1.67^{+0.15}_{-0.14}$). The best-fit lower limit of the emissivity index and breaking radius are found to be 3.3 and 4.8r$_g$, respectively. The inner disc radius during high flux is found to be in the range of 3.5-8.5 R$_g$. Table \ref{tab1} provides best-fit spectral parameters for both high and low flux energy spectra. Best-fit spectra during high and low fluxes, along with model components and the residual, are shown in the bottom left and bottom right panels of Figure \ref{fig2}, respectively. 
It may be noted that mean spectral analysis is not the main aim of the work; rather, the results can be used to understand the major spectral components that drive variabilities of low and high flux emissions.  

\subsection{Procedure of RMS and covariance spectral extraction}
Energy-dependent variability studies, particularly incorporating rms and covariance spectral analysis, are often used to track variable components across different energy bands in X-ray binaries \citet{re99,SM03,UT09}. This approach was later implemented in various analyses of AGN data \citep{ut11,mi11,ka13,ca13}. The covariance is the cross-spectral counterpart to the rms spectrum, which measures the rms amplitude of variability as a function of energy. By selecting a specific Fourier frequency range for power spectral density (PSD) integration within each energy band, a Fourier-resolved rms spectrum can be obtained \citep{re99,gi00}. Likewise, the cross-spectrum can be utilised to derive a Fourier frequency-resolved covariance spectrum, which represents the spectral shape of components correlated with the reference band. Choosing an appropriate reference band allows for the identification of spectral components that vary together within a given frequency range and those that do not. The rms and covariance spectra can be determined as follows: According to the suggestions of \citet{UT09}, in order to obtain the PDS as a function of energy, we have generated light curves corresponding to different energy bins between 0.3-9 keV. Each light curve has been divided into $M$ segments, each segment being divided into $N$ time bins. In this work, we choose values of M = 1, 3, and 5, and N = 128.0, 256.0, 512.0, and 1024.0. Values of $M$ and $N$ are chosen such that data points in each energy and time bin must have a minimum signal-to-noise ratio (ratio of count rate and the error on count rate) of 2. With this criteria, we found that using lightcurves in the following 23 energy bands (keV): 0.3-0.5, 0.5-0.7, 0.7-0.9, 0.9-1.1, 1.1-1.3, 1.3-1.5, 1.5-1.7, 1.7-1.9, 1.9-2.1, 2.1-2.3, 2.3-2.5, 2.5-2.7, 2.7-2.9, 2.9-3.1, 3.1-3.3, 3.3-3.6, 3.6-4.0, 4.0-4.5, 4.5-5.0, 5.0-6.0, 6.0-7.0, 7.0-8.0, 8.0-9.0 along with the time bin size of 128s, 256s, 512s and 1024s are suitable for the covariance spectral analysis.  Thus, we find the excess variance for each segment using \citep{UT09}:

\begin{equation}
    \label{eqn:excess-var}
    \sigma^2_{xs}=\dfrac{1}{N-1}\sum^N_{i=1}(X_i-\overline{X})^2-\overline{\sigma^2_{err}}
\end{equation}

where $X_i$ is the count rate in the $ i$th bin and $\overline{X}$ is the average count rate in the segment, and $\overline{\sigma^2_{err}}$ is the mean squared error, which, when subtracted from the variance, leaves out the excess variance. The excess variances are averaged over all the segments in each energy bin and plotted with respect to the energy that forms the rms spectra. 

We may note that at low signal-to-noise, especially at high energies, the rms spectra yield large uncertainties which are mostly consistent with zero. This can be seen in the rms spectrum in the left panel of Figure \ref{fig3} and also in Figure 2 from \citet{UT09}. In addition, it can yield negative excess variance when the expected Poisson term exceeds the measured variance, preventing further rms calculations by biasing results. This limitation can be overcome by the covariance spectrum, the cross-correlation analogue of the autocorrelation of rms spectra, which is robust to Poisson noise since uncorrelated fluctuations cancel out, and negative residuals do not affect the measurement. Therefore, we do not use the RMS spectra for any further analysis; instead, we focus on the covariance analysis. To quantify astrophysical temporal variabilities, covariance spectra are calculated in the Fourier domain over a specific range of frequencies.
We calculate the ``covariance" spectra using 
\begin{equation}
    \label{eqn:cov}
    \centering
    \sigma^2_{cov} = \dfrac{1}{N-1}\sum^N_{i=1}(X_i-\overline{X})(Y_i-\overline{Y})
\end{equation}
Where $Y_i$ is the count rate of the ${i}$th time bin in a reference band (0.3-0.5 keV here), the reference band is chosen such that the signal-to-noise ratio is very high. $\sigma^2_{cov}$ is again averaged over all the segments for each energy bin and plotted against energy to generate the covariance spectra. In regions where the signal-to-noise ratio of the other light curves is low, the count rates will be averaged to give a null contribution. Additionally, the spectra highlight regions where the signal variability from the specific energy bin is correlated with the signal variability from the reference band. For uncorrelated spectra, $\sigma^2_{cov}$ will be statistical fluctuations around zero intrinsic covariance. The covariance can be normalised using the excess variance of the reference band as follows:

\begin{equation}
    \label{eqn:norm-cov}
    \centering
    \sigma_{cov,norm}=\dfrac{\sigma^2_{cov}}{\sqrt{\sigma^2_{xs,y}}}
\end{equation}
Thus, the reference variance must not be negative. The errors in covariance \citep{UT09} are given by:

\begin{equation}
    \label{eqn:cov-err}
    Err[\sigma_{cov,norm}]=\sqrt{\dfrac{\sigma^2_{xs,x}\ \overline{\sigma^2_{err,y}} + \sigma^2_{xs,y}\ \overline{\sigma^2_{err,x}} + \overline{\sigma^2_{err,x}}\ \overline{\sigma^2_{err,y}}}{NM\sigma^2_{xs,y}}}
\end{equation}

Covariance and its errors are calculated using the above equations in 23 different energy bands provided earlier.

\subsection{Fourier-frequency resolved covariance spectra}
Covariance has been calculated in different Fourier frequency segments by varying the total length of the light curve as well as the bin size, as shown in Table \ref{tab-freq}. For calculating the lower ends of the Fourier-frequency range, the light curve (T) is divided into n segments, where n = 1, 4 and 6. Therefore, the lower limit of the Fourier frequency, $f_{low}$, is $n/T$ Hz. Higher ends of the Fourier frequency are determined by the Nyquist frequency, $f_{high}$ = 1/(2*$\delta$t) Hz, where $\delta$t is the bin size of the lightcurve (128s, 256s, 512s and 1024s in our work). Fourier frequency ranges corresponding to different bin sizes and segment lengths are provided in Table \ref{tab-freq}. Throughout the paper, we use the notation like `S4B3’ to denote that the analysis is performed over Fourier frequency ranges of 6.7-97 $\times$ 10$^{-5}$ Hz, roughly corresponding to time scales between 512 sec and 15000 sec.

\begin{table}
\centering
\begin{tabular}{c|cccc}
\hline

\textbf{Seg no.} & 128 sec & 256 sec & 512 sec & 1024 sec\\
 & B1 & B2 & B3 & B4 \\
\hline
S1 (T) & 1.1-390 & 1.1-195 & 1.1-97 & 1.1-48  \\

S4 (T/4) & 6.7-390 & 6.7-195 & 6.7-97 & 6.7-48  \\

S6 (T/6) & 9.9-390 & 9.9-195 & 9.9-97 & 9.9-48  \\
\hline
\end{tabular}
\caption{Table of Fourier frequency ranges (in the unit of 10$^{-5}$ Hz) used to extract covariance spectra. B1, B2, B3 and B4 stand for bin sizes of 128sec, 256sec, 512sec and 1024sec, respectively, while S1, S4 and S6 imply 1, 4, and 6 lightcurve segments T = maximum lightcurve length of nearly 65 ksec. A notation `S4B2' in the text implies the analysis is carried out in the Fourier frequency ranges 6.7-195 $\times$ 10$^{-5}$ Hz or time ranges roughly corresponding to 256 sec-15000sec.}
\label{tab-freq}
\end{table}

Therefore, energy-dependent covariance spectra are calculated for a range of Fourier-frequencies with a minimum of 1.5 $\times$ 10$^{-5}$ Hz to the maximum value of 390 $\times$ 10$^{-5}$ Hz. Our choice of Fourier frequency ranges is decided by two factors: (1) We ensure that the number of bins in each light curve is sufficiently high to calculate covariance spectra. The minimum and maximum number of bins in the light curve varies between 13 and 507, depending on segment length and bin size. (2) We also make sure that the signal-to-noise ratio (SNR) in each bin is sufficiently high (at least 5) in all energy bands. This is particularly important in the higher energy bands where SNR drops significantly.

The RMS and covariance spectra for the S4B3 segment are shown in the top and bottom panels of Figure \ref{fig3}, during low and high fluxes, respectively. We may note that although energy-dependent variability of rms and covariance spectra are similar, errors on rms are larger and unbounded, particularly in high energies ($>$ 2 keV) during both flux states. A similar problem is observed during the RMS and covariance spectral analysis of BHXRBs \citep{UT09}. However, covariance errors are bounded in all energies. Therefore, we use covariance spectra throughout the rest of the work.
In order to check variations in the covariance spectra as a function of Fourier frequencies, we fitted 0.5-9 keV covariance spectra with a single powerlaw with the powerlaw index fixed to that obtained from mean spectral analysis, while allowing normalisation to vary in all spectra. We show the data to model ratio for S1B4, S4B3 and S6B1 segments, respectively, during low flux (top panel of Figure \ref{fig4}) and high flux (bottom panel of Figure \ref{fig4}). We may note that (1) covariance spectra cannot be fitted with a single powerlaw at any Fourier frequencies, (2) there is a significant difference in ratio values at different photon energies for three different Fourier frequencies, particularly below 2 keV. Such an observation implies that Covariance spectra at different Fourier frequencies do change during the soft excess in NGC 4593.

\subsection{Covariance spectral modelling and results}
The covariance spectrum traces variability between energy bands simultaneously over a chosen timescale. Such a correlated variability can be interpreted as arising from different spectral components. For NGC 4593, mean spectral fitting shows X-ray emission dominated by the coronal power-law continuum, with a secondary contribution from reflection and absorption. If variability is due to continuum normalisation, the covariance matches the mean (same slope); if due to photon index changes, it may be harder or softer. Differently varying reflection or absorption causes deviations (e.g., weaker Fe line or soft excess). Thus, comparing covariance and mean spectra helps identify whether a single dominant coronal component drives variability or if additional constant/weakly variable features contribute. 
However, it may be noted that the covariance spectra can be modelled using the same spectral components as the time-averaged spectra only when the dominant mode of variability corresponds to changes in the normalisation of those components. If the variability instead involves changes in spectral shape, even as simple as a variation in the powerlaw slope, higher-order terms are expected to dominate the covariance spectrum. For example, the hump-shaped residuals observed around 1 keV in Figure \ref{fig4} may indicate a change in the slope of the primary power-law component, which needs to be verified \citep{ta25}.

Motivated by this and following similar arguments presented in \citet{UT09}, each covariance spectrum is modelled using a combination of a power-law continuum and a reflection continuum in the presence of an absorption component if needed.

Therefore, best-fit covariance spectra are obtained by \textsc{zpowerlw+xillver} or \textsc{zpcfabs*(zpowelw+xillver)} model at various Fourier frequencies. We have used {\textsc zpcfabs} as the absorption model for covariance spectra since it is simpler than {\textsc zxipcf} (has one less parameter).  Figure \ref{fig5} shows best-fit covariance spectra along with residuals during low and high fluxes, respectively. The rest of the plots are shown in the Appendix \ref{sec:append}. Tables \ref{tab2} provide best-fit parameters for covariance spectral fitting for one set of Fourier frequencies during low flux, while the same values during high flux are provided in Table \ref{tab3}, respectively. Remaining Tables for best-fit parameters are provided in the Appendix \ref{sec:append}. From fitted parameters, we may note that absorption is only required statistically during low Fourier frequency regimes when the 65 ksec light curve has been considered to calculate the covariance. Modelling covariance at higher Fourier frequencies does not require an absorption. Such modelling behaviour is expected since low-frequency covariance spectra will be more similar to mean-energy spectra, where the mean-spectral modelling requires the presence of significant absorption.  
We have computed 68\%, 90\% and 95\% contours for the normalisation of powerlaw and reflection model components from best-fit covariance spectra during low and high fluxes and shown in Figure \ref{fig-cont}. During the high Fourier frequency regime (8.8-390 $\times$ 10$^{-5}$ Hz), both powerlaw and reflection normalisations are well constrained during low-flux observations. However, during high flux, powerlaw normalisation is small and unconstrained, while the reflection norm is well constrained, at least up to the 95\% confidence level. 
During low Fourier frequencies, contours are constrained at least up to 90\% confidence for both powerlaw and reflection normalisations.

Comparing best-fit parameters in different Fourier frequencies during low and high flux covariance spectra, we may note that the effect of absorption becomes insignificant at higher Fourier frequencies. In all Tables, absorption is absent at the highest frequencies: 9.9-390, 9.9-195, 9.9-97 and 9.9-48 $\times$ 10$^{-5}$ Hz. This is expected as higher frequency covariance spectra are expected to be less affected by absorption variability since absorption can only vary on a timescale of days to weeks, which is much slower than the timescale considered here for covariance analysis. Also, the powerlaw index of the direct continuum tends to be flatter at higher Fourier frequencies. For example, the best-fit power-law indices of the powerlaw continuum during low flux are found to be $2.35 ^{+0.06}_{-0.09}$, $2.13^{+0.29}_{-0.23}$ and $1.77^{+0.06}_{-0.17}$ when the lower limit of Fourier frequencies start at 2.1-, 6.7- and 9.9- $\times$ 10$^{-5}$ Hz respectively (see Table \ref{tab2}).
Instead of studying individual parameters, we have calculated the overall contribution of the normalisation of each continuum component to the total normalisation and plotted them (in percentage) during low and high fluxes as a function of Fourier frequencies in the left and right panels of Figure \ref{fig7}. During low flux, variabilities in all Fourier frequencies are dominated by the power-law continuum over reflection. However, the scenario is different during high flux: the reflection continuum mostly dominates the variabilities, particularly at high Fourier frequencies. Such an observation implies that mean spectra are more similar to reflection spectra than powerlaw spectra during high flux, but during low flux, powerlaw spectra drive the variabilities.

\begin{table}
\centering
\renewcommand{\arraystretch}{1.5}
\begin{tabular}{cccc}
\hline
\thead{{\bf Model} \\ {\bf components}} & \thead{{\bf Parameters}} & \thead{{\bf High} \\ {\bf flux}} & \thead{{\bf Low} \\ {\bf flux}}  \\
\hline
\hline
{\tt zxipcf$_{1}$} & \thead{$N_H$ \\ ($10^{22}\ cm^{-2}$)} & $0.76^{+0.13}_{-0.14}$ & $0.19^{+0.03}_{-0.01}$ \\

{\tt zxipcf$_{1}$} & \thead{$\log_{\xi}$ \\ ($ergs\ cm^{-2}\ s^{-1}$)} & $1.09^{+0.12}_{-0.28}$ & $2.73^{+0.02}_{-0.02}$ \\

{\tt zxipcf$_{1}$} & $CvrFract$ & $0.26^{+0.02}_{-0.01}$ & $<0.6$ \\

{\tt zxipcf$_{2}$} & \thead{$N_H$ \\ ($10^{22}\ cm^{2}$)} & $0.55^{+0.24}_{-0.18}$ & $0.83^{+0.03}_{-0.13}$ \\

{\tt zxipcf$_{2}$} & \thead{$\log_{\xi}$ \\ ($ergs\ cm^{-2}\ s^{-1}$)} & $2.72^{+0.07}_{-0.09}$ & $0.86^{+0.02}_{-0.01}$ \\

{\tt zxipcf$_{2}$} & F$_{cvr}$ & $0.46^{+0.10}_{-0.12}$ & $0.33^{+0.01}_{-0.01}$ \\

{\tt kdblur} & I$_1$ & $>$3.3 & -- \\

{\tt kdblur} & R$_{in}$ (R$_g$) & $5.4^{+3.1}_{-1.9}$ & -- \\

{\tt kdblur} & R$_{break1}$ & $>$4.8r$_g$ & -- \\

{\tt kdblur} & I$_{2}$ & $>$5.48 & -- \\

{\tt kdblur} & R$_{break2}$ & 20r$_g$ (f) & -- \\

{\tt kdblur} & I$_{3}$ & 10 (f) & -- \\


{\tt xillver$_{sx}$} & $\Gamma_{sx}$ & =$\Gamma_{PL}$ & -- \\

{\tt xillver$_{sx}$} & \thead{$\log_{\xi}$\\ ($ergs\ cm^{-2}\ s^{-1}$)} & $2.67^{+0.15}_{-0.14}$ & --\\

{\tt xillver$_{sx}$} & $A_{fe}$ & $<0.56$ & -- \\

{\tt xillver$_{sx}$}  & $E_{cut}$ & 300 (f) & -- \\

{\tt xillver$_{sx}$} & \thead{$norm$ \\ ($\times10^{-4}$)} & $3.74^{+0.41}_{-0.25}$ & -- \\

{\tt xillver} & $\Gamma$ & =$\Gamma_{PL}$ & =$\Gamma_{PL}$ \\

{\tt xillver} &  \thead{$\log_{\xi}$\\ ($ergs\ cm^{-2}\ s^{-1}$)} & $<0.7$ & $0.36^{+0.24}_{-0.11}$ \\

{\tt xillver} & $A_{fe}$ & $<$0.87 & $<$0.68 \\

{\tt xillver} & $E_{cut}$ & $203(f)$ & $<389$ \\

{\tt xillver} & \thead{$norm$ \\ ($\times10^{-4}$)} & $0.22^{+0.26}_{-0.06}$ & $2.61^{+0.13}_{-0.12}$ \\

{\tt powerlaw} & $\Gamma_{PL}$ & $1.97^{+0.02}_{-0.01}$ & $2.01^{+0.03}_{-0.03}$ \\

{\tt powerlaw} & \thead{$norm$ \\ ($\times10^{-2}$)} & $1.45^{+0.02}_{-0.01}$ & $0.42^{+0.03}_{-0.01}$ \\

{\tt gabs} & $line_{E}\ (keV)$ & $6.04^{+0.03}_{-0.04}$ & $6.08^{+0.02}_{-0.02}$ \\

{\tt gabs} & $\sigma$ & $2.17^{+0.14}_{-0.18}$ & $1.42^{+0.02}_{-0.02}$ \\

{\tt gabs} & $strength$ & $<$0.1 & $<$0.1 \\

Abs Flux & (10$^{-11}$ cgs) & $8.48^{+0.02}_{-0.03}$ & $3.62^{+0.03}_{-0.02}$ \\
\hline
$\chi^2/dof$ & & \thead{$177/141$ \\ } & \thead{$171/147$ \\ } \\
\hline

\end{tabular}
\caption{Best-fitting model parameters for the {\tt EPIC-PN} 0.3-9 keV spectrum of NGC 4593 for high flux and low flux observations.}
\label{tab1}
\end{table}

\begin{figure}
\centering
\includegraphics[scale=0.3,angle=-90]{rms-cov-low.ps}
\includegraphics[scale=0.3,angle=-90]{rms-cov-high.ps}
\caption{Photon-energy dependent rms (triangle) and covariance (circle) spectra extracted during the low and high fluxes observation for the S4B3 segment (see Table \ref{tab-freq}) are shown in the top and bottom panels, respectively. During both flux states, unbounded errors are observed in rms spectral bins mostly above 2 keV; however, covariance errors are constrained at all energies. }
\label{fig3}
\end{figure}

\begin{figure}
\centering
\includegraphics[scale=0.32,angle=-90]{ratio-2016.ps}
\includegraphics[scale=0.32,angle=-90]{ratio-2002.ps}
\caption{Comparison of covariance spectra at different Fourier frequencies when fitted with a powerlaw: Top panel shows data to powerlaw fitted model ratio for 0.3-9 keV covariance spectra extracted in the frequency range 1.1-48 (grey dots; seg S1B4), 6.7-97 (red triangles; seg S4B3) and 9.9-390 (black stars; seg S6B1) $\times$ 10$^{-5}$ Hz during low flux. A significant departure from the ratio=1 line at different energies and differences in ratio values can be observed at different Fourier frequencies. The bottom panel shows the same during the high flux. }
\label{fig4}
\end{figure}

\begin{figure*}
\includegraphics[scale=0.32,angle=-90]{cov-spec4-fig1.ps}
\includegraphics[scale=0.32,angle=-90]{cov-spec4-fig2.ps} \\
\includegraphics[scale=0.32,angle=-90]{cov-spec4-fig3.ps}
\includegraphics[scale=0.32,angle=-90]{cov-spec4-fig5.ps}\\
\caption{Modelling of covariance spectra: best-fit covariance energy spectra fitted with a combination of powerlaw and reflection continuum are shown along with model components and residuals in the Fourier frequency (in the unit of 10$^{-5}$ Hz) range of  6.7-390 (top left) and 6.7-195 (top right) during the low flux while the same during the high flux are shown in the bottom left and bottom right panels respectively.}
\label{fig5}
\end{figure*}

\begin{table*}
\centering
\renewcommand{\arraystretch}{1.5}
\begin{tabular}{ccccc}
\hline
\thead{{\bf Model}} & & Fourier & frequency & segments \\ 
{\bf components} & 
\thead{{\bf Parameters}} & S1B2 & S4B2 & S6B2 \\
\hline
{\tt zpcfabs} & \thead{$N_H$ } & $0.42^{+0.12}_{-0.14}$ & $0.57^{+0.23}_{-0.22}$ & $--$ \\

{\tt zpcfabs} & $F_{cvr}$ & $0.62^{+0.08}_{-0.11}$ & $0.74^{+0.09}_{-0.10}$ & $--$ \\

{\tt xillver} & $\Gamma_{xill}$ & =$\Gamma_{PL}$ & =$\Gamma_{PL}$  & =$\Gamma_{PL}$ \\

{\tt xillver} & $A_{fe}$ & $4.47^{+0.51}_{-1.79}$ & $4.38^{+0.14}_{-1.22}$ & $4.7^{+0.1}_{-0.9}$ \\

{\tt xillver} & \thead{$\log_{\xi}$} & $2.94^{+0.15}_{-0.11}$ & $2.82^{+0.23}_{-0.11}$ & $2.83^{+0.09}_{-0.07}$ \\

{\tt xillver} & \thead{$norm$  ($\times10^{-2}$)} & $2.19^{+0.08}_{-0.06}$ & $0.39^{+0.05}_{-0.01}$  & $0.19^{+0.04}_{-0.02}$\\

{\tt powerlaw} & $\Gamma_{PL}$ & $2.35 ^{+0.06}_{-0.09}$ & $2.13^{+0.29}_{-0.23}$  &$1.77^{+0.06}_{-0.17}$ \\

{\tt powerlaw} & \thead{$norm$  ($\times10^{-2}$)} & $2.9^{+0.020}_{0.027}$ & $0.97^{+0.01}_{-0.009}$ & $0.8^{+0.006}_{-0.003}$ \\

\hline
$\chi^2/dof$ & & \thead{$15.71/16$} & \thead{$16.58/16$ } &\thead{ $18.01/18$ } \\
\hline
\end{tabular}

\caption{Bestfit parameters for 0.5-9.0 keV covariance photon energy spectra extracted for S1B1, S4B1 and S6B1 Fourier segments (see Table \ref{tab-freq}) during low flux are provided along with their models and 1$\sigma$ errors. $F_{cvr}$ is the covering fraction. The unit of N$_H$, Hydrogen column density is $10^{22}\ cm^{-2}$, $\xi$ is the ionization parameter with the unit of $ergs\ cm^{-2}\ s^{-1}$. The high energy cutoff, disc inclination angle and reflection fraction are kept fixed at 300 keV, 56$^o$ and -1.0, respectively, in all fittings. }
\label{tab2}
\end{table*}

\begin{table*}
\centering
\renewcommand{\arraystretch}{1.5}
\begin{tabular}{ccccc}
\hline
\thead{{\bf Model}} & & Fourier & frequency & segments \\ 
{\bf components} & 
\thead{{\bf Parameters}} & S1B1 & S4B1 & S6B1 \\
\hline
{\tt zpcfabs} & \thead{$N_H$} & $0.43^{+0.13}_{-0.04}$ & -- & -- \\

{\tt zpcfabs} & $F_{cvr}$ & $0.71^{+0.05}_{-0.06}$ & -- & -- \\

{\tt xillver} & $\Gamma_{xill}$ & =$\Gamma_{PL}$ & =$\Gamma_{PL}$  & =$\Gamma_{PL}$ \\
{\tt xillver} & $A_{fe}$ & $3.13^{+0.82}_{-0.11}$ & $4.57^{+0.72}_{-1.10}$ & $2.88^{+1.27}_{-0.96}$ \\

{\tt xillver} & \thead{$\log_{\xi}$} & $3.79^{+0.11}_{-0.20}$ & $3.67^{+0.14}_{-0.09}$ & $3.04^{+0.0}_{-0.0}$ \\

{\tt xillver} & \thead{$norm$  ($\times10^{-2}$)} & $0.95^{+0.06}_{-0.04}$ & $0.16^{+0.02}_{-0.01}$  & $2.21^{+0.02}_{-0.3}$\\

{\tt powerlaw} & $\Gamma$ & $2.88^{+0.22}_{-0.24}$ & $2.39^{+0.37}_{-0.54}$  &$2.02^{+0.16}_{-0.10}$ \\

{\tt powerlaw} & \thead{$norm$  ($\times10^{-2}$)} & $0.12^{+0.04}_{-0.03}$ & $2.19^{+0.02}_{-0.03}$ & $0.22^{+0.03}_{-0.01}$ \\

\hline
$\chi^2/dof$ & & $14.90/16$ & $14.50/18$ & $14.13/18$ \\
\hline
\end{tabular}
\caption{Bestfit parameters for 0.5-9.0 keV covariance photon energy spectra extracted for S1B1, S4B1 and S6B1 Fourier segments (see Table \ref{tab-freq}) during high flux are provided along with their models and 1$\sigma$ errors. $F_{cvr}$ is the covering fraction. The unit of N$_H$, Hydrogen column density is $10^{22}\ cm^{-2}$, $\xi$ is the ionization parameter with the unit of $ergs\ cm^{-2}\ s^{-1}$. The high energy cutoff, disc inclination angle and reflection fraction are kept fixed at 300 keV, 56$^o$ and -1.0, respectively, in all fittings. }
\label{tab3}
\end{table*}

\begin{figure*}
\centering
\includegraphics[scale=0.3,angle=-90]{lag-2016-percent.ps}
\includegraphics[scale=0.3,angle=-90]{lag-2002-percent.ps}
\caption{Plot of direct and reflection continuum models contribution (in percentage to the total contribution) to the covariance spectra as a function of Fourier frequencies during low flux (left panel) and high flux (right panel) observations.}
\label{fig7}
\end{figure*}

\begin{figure*}
\centering
\includegraphics[scale=0.3,angle=-90]{cont-2016.ps}
\includegraphics[scale=0.3,angle=-90]{cont-2002.ps}
\includegraphics[scale=0.3,angle=-90]{cont-2016-1024.ps}
\includegraphics[scale=0.3,angle=-90]{cont-2002-1024.ps}
\caption{Plot of contours (68\% (red), 90\% (green) and 95\% (blue)) for direct and reflection continuum model normalisations are shown from best-fit covariance spectra during low (top left) and high (top right) flux for S6B1 Fourier segment. The same is true during the S1B4 segment, which is shown for low (bottom left) and high (bottom right) flux observations.}
\label{fig-cont}
\end{figure*}

\subsection{Power density and time-lag spectral analysis and results}
To understand the strength of variabilities, we have computed power density spectra (PDS) from the clean, continuous light curves during high- and low-flux observations, and shown in the top and bottom panels of Figure \ref{fig8}, respectively. PDS are extracted for 0.3-1 keV (soft band) and 1-5 keV (hard band) with a time bin size of 128 sec, so that the Nyquist frequency is 3.9 mHz. PDS are rms-normalized, and Poisson noise was not subtracted. A logarithmic binning of -1.5 is used to present PDS in soft and hard bands. 
PDS from both observations show a fractional rms power significantly higher than the expected Poisson noise level, in the Fourier frequency range of 0.02-1 mHz. A moderate variability is observed up to 3.9 mHz in the 1-5 keV hard band PDS during high flux. Therefore, we have confined our analysis to below 3.9 mHz, however, results from lag and covariance analysis above 1 mHz may be consistent with zero. 
We may note that the rms power is significant during low flux below 5 $\times$ 10$^{-5}$ Hz. Also, at any given Fourier frequency, the rms power during low flux is higher than that of high flux (e.g., at 0.1 mHz, low-flux rms power is nearly a factor of 2 higher than that of high flux in 1-5 keV), indicating a more compact geometry of corona during low flux than that during high flux. 

We estimated the phase/time lag between two time series in different energy bands using standard analysis methods outlined in \citet{be00} and \citet{no99}. We have chosen two clean, background-subtracted light curves: 0.3-1 keV, denoted as \( s(t) \) (soft band) and 1-5 keV \( h(t) \) (hard band), obtained simultaneously during low and high flux observations. These bands effectively represent the spectral regime of soft excess and the illuminating X-ray source (i.e., continuum), though the exact energy ranges for each source may vary slightly. These light curves consist of \( N \) (= 493 in our case) equidistant observations with a sampling period \( t_{\text{bin}} \) ( = 128 sec in our case).
For each Fourier frequency \( f_j = \frac{j}{N t_{\text{bin}}} \), where \( j = 0, 1, \ldots, \left\lfloor \frac{N}{2} \right\rfloor \) (for even \( N \)) or \( \left\lfloor \frac{N-1}{2} \right\rfloor \) (for odd \( N \)), we compute the cross-spectrum between the two light curves.
The cross-spectrum \( F_{sh}(f_j) \) at frequency \( f_j \) represents the Fourier transform of the cross-correlation function between \( s(t) \) and \( h(t) \). From the cross-spectrum \( F_{sh}(f_j) \), we can derive the time-lag spectrum. The time lag \( \tau_j \) at frequency \( f_j \) is given by:
   \[ \tau_j = \frac{\text{Arg}(F_{sh}(f_j))}{2 \pi f_j} \]
   where \( \text{Arg}(F_{sh}(f_j)) \) is the phase of the cross-spectrum \( F_{sh}(f_j) \).
  The standard deviation \( \text{std}\{\tau(f_{\text{bin}}, i)\} \) for each phase-lag estimate is calculated using equations 16 and 17 from \citet{no99}. These equations provide a method to quantify the uncertainty in the time-lag estimates at different Fourier frequencies. The coherence \( \gamma_{sh}(f_j) \) between the soft \( s(t) \) and hard \( h(t) \) light curves is estimated from the cross-spectrum as a function of Fourier frequency \( f_j \). 
Coherence is defined as: \[ \gamma_{sh}(f_j) = \frac{|F_{sh}(f_j)|^2}{P_s(f_j)P_h(f_j)} \] where \( |F_{sh}(f_j)|^2 \) is the magnitude squared of the cross-spectrum, and \( P_s(f_j) \) and \( P_h(f_j) \) are the power spectral densities of the soft and hard light curves, respectively.
Coherence takes values between 0 and 1, indicating the degree of linear correlation between the soft and hard light curves at a given Fourier frequency. High coherence values (close to 1) indicate a strong linear correlation between the phases of \( s(t) \) and \( h(t) \) at a specific Fourier frequency, suggesting acceptable time-lag estimates while low coherence values (close to 0) imply weak or no correlation between the phases \( \phi_S(f_j) \) and \( \phi_H(f_j) \). 

\begin{figure}
\centering
\includegraphics[scale=0.34,angle=-90]{pds-high.ps}
\includegraphics[scale=0.34,angle=-90]{pds-low.ps}
\caption{Power density spectra extracted for soft band (0.3-1 keV; shown by dots) and hard band (1-5 keV; shown by crosses) are shown for high flux (top panel) and the low flux (bottom panel) observations, respectively. Dashed horizontal lines in both panels show the estimated Poisson noise level during soft (grey dashed line) and hard (black dashed line), respectively.}
\label{fig8}
\end{figure}

\begin{figure}
\centering
\includegraphics[scale=0.3,angle=-90]{high-coherence.ps} \\
\includegraphics[scale=0.3,angle=-90]{low-coherence.ps} \\
\includegraphics[scale=0.3,angle=-90]{lag-high-low-4593.ps}
\caption{Coherence between 0.3-1 keV and 1-5 keV is shown as a function of Fourier frequency during high flux (top panel) and low flux (middle panel), respectively. Time lag spectra (lag between 0.3-1 keV and 1-5 keV) during low and high fluxes are shown in the bottom panel.}
\label{fig9b}
\end{figure}

\begin{figure*}
\centering
\includegraphics[scale=0.40,angle=0]{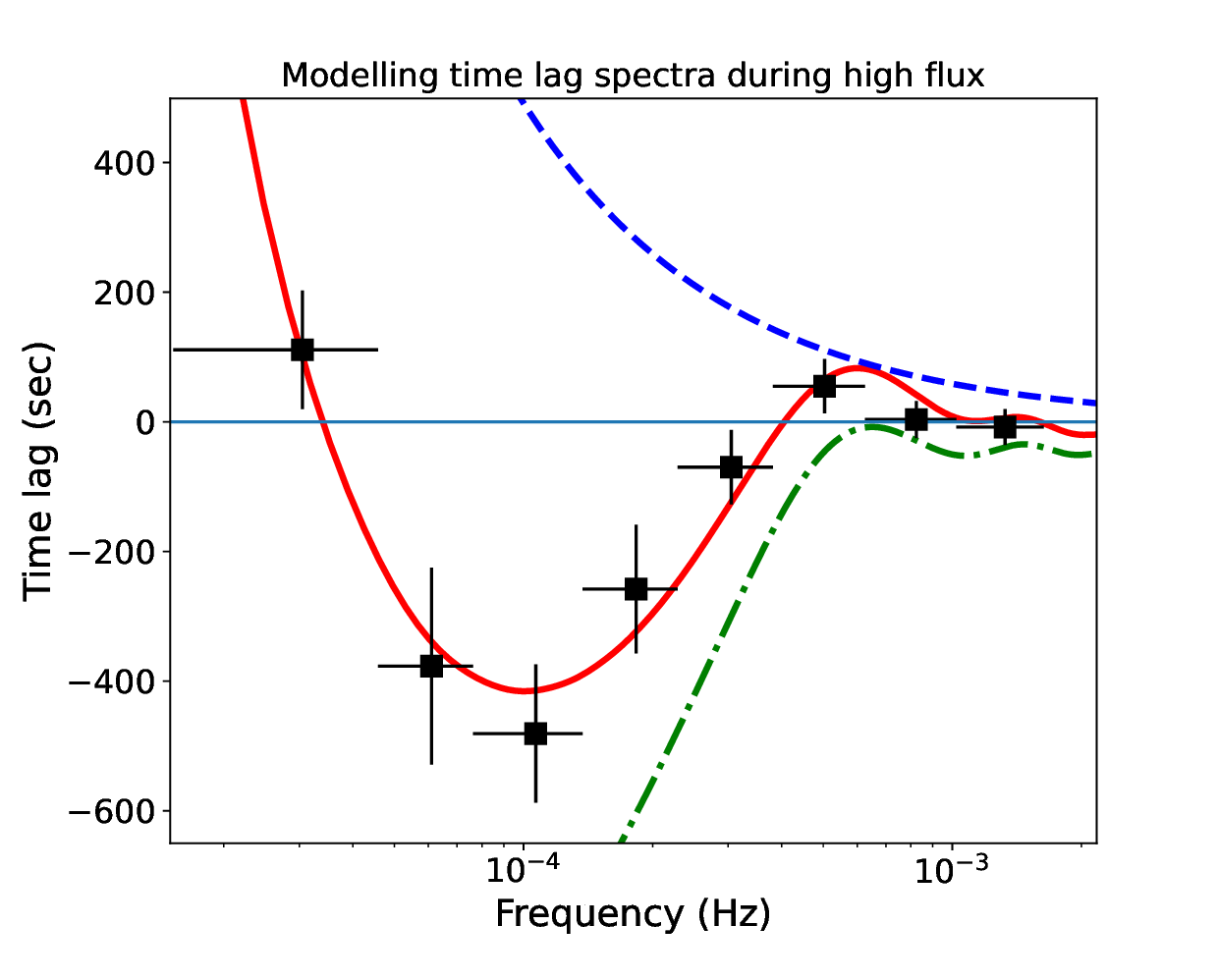}
\includegraphics[scale=0.40,angle=0]{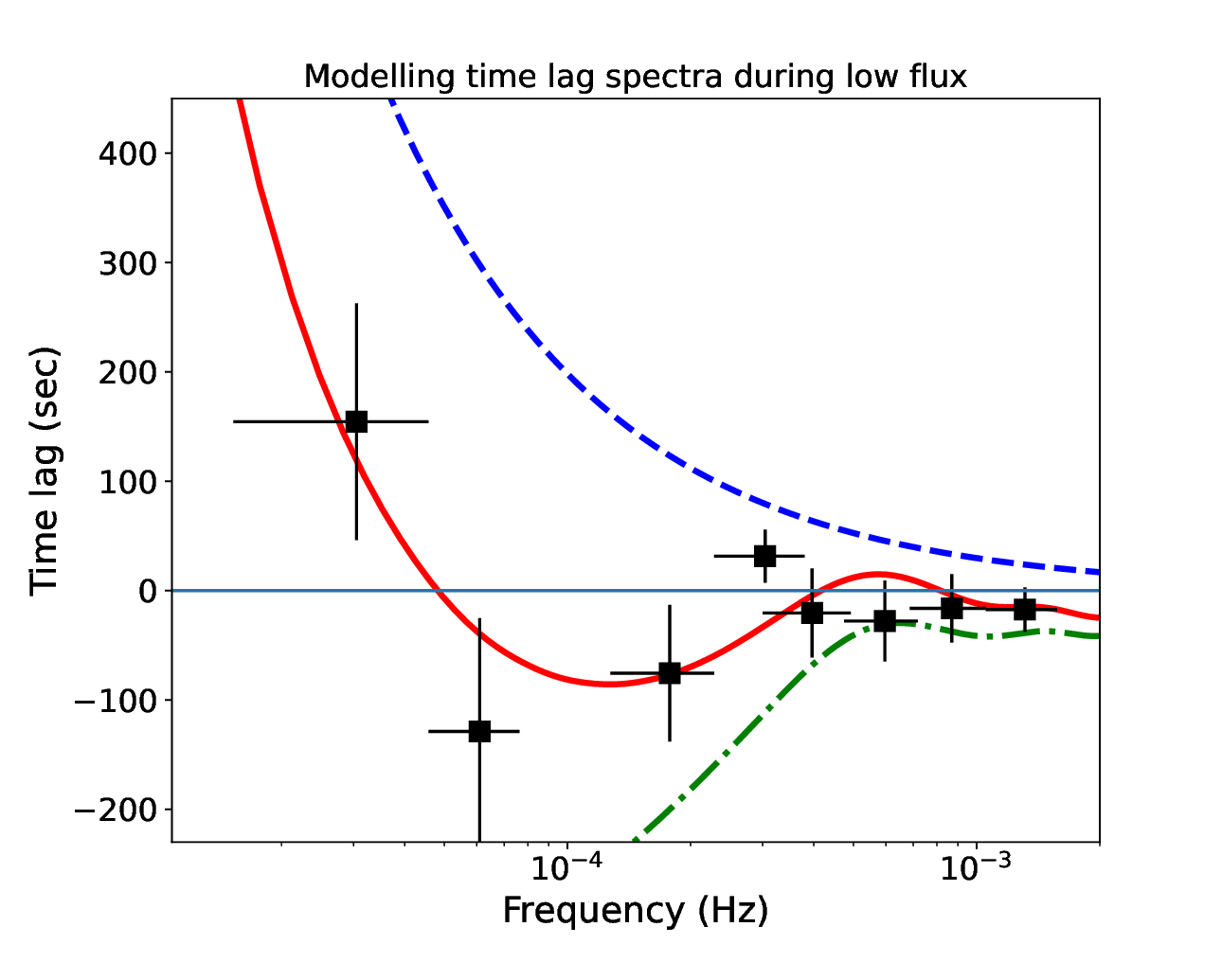}
\caption{time-lag spectra between 0.3-1 keV and 1-5 keV are shown as a function of Fourier frequency during high flux (left panel) and low flux (right panel), respectively. Positive values indicate hard photons lagging behind soft photons, while negative values indicate the opposite behaviour (reverberation delay). Time lag spectra are modelled with a combination of the accretion disc propagation fluctuation model (shown in a dotted line) and a General Relativistic reverberation model based on lamp-post geometric configurations (shown in a dot-dashed line). The resultant fit is shown in a solid line. }
\label{fig10}
\end{figure*}

Following the above, we have calculated the coherence between the soft (0.3-1 keV) and hard (1-5 keV) bands as a function of Fourier frequencies and have shown them in the left and right panels of Figure \ref{fig9b} during high and low fluxes, respectively. Poisson noise corrections are applied to coherence calculations.
We may note that coherence is high ($\geq$ 0.3) below 1.5 mHz, however, it falls off quickly to 0.15 or below at high frequencies. A coherence of 0.3 or higher is consistent with the lag analysis prescription given by \citet{ep16}, where acceptable coherence depends on the number of segments used to calculate the cross-spectrum, and coherence of 0.3 or higher is significant for a number of lightcurve segments of 15 or higher. This prescription is followed in the current analysis.
We estimate the time-lag spectra down to (1-25) $\times$ 10$^{-4}$ Hz, but for the lag spectral fitting, we consider only the time-lag estimates for which the coherence is greater than 0.3, corresponding to a physically meaningful phase correlation.
 Fourier-frequency-dependent lag spectra between the soft (0.3-1 keV) and hard (1-5 keV) bands are shown during high and low flux observations in the bottom panel of Figure \ref{fig9b}. Positive values of time lag imply that hard photons are lagged behind soft, reprocessed photons, while negative lag implies that soft photons are delayed compared to hard photons. We use logarithmic frequency binning, with each bin spanning a factor of 1.5 to improve the quality of the estimate while decreasing sampling fluctuations. We assume uncertainties on lag measurements in different Fourier frequencies are normally distributed \citep{va97,no99,ut14}.
 
Left and right panels of Figure \ref{fig10} shows high and low flux lag spectra which are modelled using two components: (1) A power law-like component (with two variables) that depicts the long, positive lag due to propagation fluctuations in the accretion disc \citep{lyu97} (shown by dashed line) (2) the short, negative lag are modelled with the reverberation lag model (with two variables) \footnote{\url{https://projects.asu.cas.cz/stronggravity/kynreverb}} assuming lamp-post geometry \citep{emm14,dov04} (shown by dot-dashed line). We have generated a set of theoretical lag spectral models by varying the coronal height parameter from 1.2 to 15 R$_g$ and disc ionisation parameter (log $\xi$) from -3 to +4 while keeping all other parameters fixed at values either obtained from mean spectral analysis or using published literature. Therefore, we used four variables to fit the lag spectra, which consist of 8 data points. Although the number of data points in the lag spectrum to fit is 8, $\chi^2$ values for fitting the lag model are 7.8 and 8.6 for high and low flux lag spectra, respectively.

Errors on fitted parameters are calculated using Markov Chain Monte Carlo simulations of the fitted spectra using uncertainties on lag values. During fitting, we kept the power law index and variability amplitude as free parameters for the fluctuation model, while the height of the corona and disc ionisation parameters were kept free to vary in the reverberation model. Best-fit lag spectral parameters show that the 95\% lower limit of coronal height or size is at least 12.2 R$_g$ during high flux, while the upper limit of the same for low flux is 7.1 R$_g$. During the transition from high to low flux, the disk ionisation parameter (log $\xi$) is also found to decrease from 2.6$\pm$1.4 to $<$0.3. All uncertainties and limits on estimated parameters are quoted with 95\% confidence. This is also consistent with the mean spectral analysis shown in Table \ref{tab1}. The lag amplitude of the propagating fluctuation is found to be higher during low flux compared to high flux.

\section{Discussions}\label{sec4}
In this work, we have studied two long \xmm{} observations of NGC 4593 taken in 2002 and 2016. While the average X-ray count rate in 0.3-10 keV during the 2002 observation was 18$\pm$ 0.3 counts/sec, the same for the 2016 observation was 5.3 $\pm$ 0.1 counts/sec. Therefore, in both observations, the X-ray count rate differs by a factor of $\sim$3.5, and the difference is significantly consistent throughout the last 65 ks of both observations. Additionally, short-term X-ray flickering is predominantly visible in the lightcurve of high X-ray flux (right panel of Figure \ref{fig1}). To understand the origin of variability and its link to the disc-corona geometry, we have performed mean, Fourier-frequency dependent covariance and time-lag spectral analysis in a self-consistent manner.
Mean spectral analysis shows that a soft excess component below 2 keV is present during high-flux observation, which can be modelled using a blurred reflection model. This component is either absent during the low flux or not significant enough to appear as a separate component in the spectrum.  
Interestingly, during the first 20 ksec of both observations when average count rates overlaps with each other, mean spectra during low and high fluxes overlaps exactly within their uncertainties.
The mean spectral analysis, presented in Table \ref{tab1}, indicates that two ionized, partially covered absorption model components are required to understand the absorption during both states; however, the absorption is not significantly different from each other: the absorption column densities (in the unit of 10$^{22}$ cm$^{-2}$) are 0.76$^{+0.13}_{-0.14}$ and 0.55$^{+0.24}_{-0.18}$, for both components during high flux, while the same during the low flux are 0.19$^{+0.03}_{-0.01}$ and 0.83$^{+0.03}_{-0.13}$ respectively. Similarly, the combined ionisation parameter due to both absorber components is similar during both high- and low-flux. Therefore, we can safely assume that the change in flux is not due to the change in the local absorbing medium, like nearby BLR clouds or warm absorbers. Hence, the change in flux is possibly due to changes in continuum and/or reprocessed emissions only.
A different behaviour in the flux-dependent soft excess is observed in Ark 564 by \citet{yu25}. Their flux-resolved spectral analysis shows that the soft excess maintains a nearly invariant shape, with variability dominated primarily by changes in normalisation rather than spectral form. This indicates that the physical origin of the soft excess in Ark 564 remains stable over a wide range of luminosities, favouring scenarios in which it arises from a less-variable emission component, such as warm Comptonization or reprocessed disc emission, which is closely coupled to the accretion flow.

\subsection{Possible disc-corona geometry during different flux states}

Table \ref{tab1} shows that the continuum during the mean spectrum during low flux can be modelled using a combination of hard X-ray emitting powerlaw from the corona and an unblurred reflection component, possibly originating from a truncated inner disc, distant reflection or both. Narrow K$\alpha$ and K$\beta$ lines observed in the powerlaw-fitted residual spectrum (top right panel of Figure \ref{fig2}) also indicate that the relativistic reflection is not significant during low flux. The detection of photoionized emission line with variable equivalent width and flux implies that Fe K line complexes may originate from an irradiated accretion disc rather than distant material. 
During the high flux, an additional soft excess component is required to model the spectrum below 2 keV. Assuming the soft excess as part of the smeared reflection continuum, an additional blurred, non-relativistic reflection model is used, which can sufficiently describe the observed excess. 
To understand the contribution of blurred and unblurred reflection at the Fe K$\alpha$ complex during high flux, we calculated the flux in 6.0-7.5 keV for blurred and unblurred components using \textsc{cflux} in \textsc{XSpec}. We noticed nearly similar flux values (1.8$^{+0.3}_{-0.2}$ and 2.2$^{+0.1}_{-0.2}$ $\times$ 10$^{-12}$ ergs s$^{-1}$ cm$^{-2}$) during both components. Due to high disc ionisation parameter (log $\xi$ = 2.67$^{+0.16}_{-0.14}$) and low Iron abundance ($A_{Fe} < 0.56$), the blurred, relativistic reflection is possibly weakened around the Fe k$\alpha$ complex (6-7.5 keV).
Fourier-frequency dependent covariance spectral analysis shows that the spectral variability during low flux is dominated by a direct continuum emission rather than a reflected continuum (left panel of Figure \ref{fig7}). It may be noted that the time scale in which the variability is studied ranges from 128 sec to 18 hours. This range is still significantly smaller than the typical timescale expected for the reverberation from the broad- or narrow-line regions ($\sim$3-4 days or longer). Therefore, observed variability, although low (30\% or less), is either due to reverberation from the accretion disc or extended corona or a disc-corona coupled system.
The modelling of the time-lag spectra between the soft (0.3-1 keV) and hard (1-5 keV) bands as a function of Fourier frequencies shows that the upper limit of coronal height or spatial extent is 3.3 R$_g$ or less during the low-flux state, while the same during the high-flux state is 7.2 R$_g$ or higher. Additionally, disc ionisation is also significantly lower during low flux than in the high flux state. Mean spectral analysis also supports the latter.

Now, assuming the mass of the black hole as 7 $\times$ 10$^{6}$ M$\odot$ \citep{mc18}, the gravitational radius, R$_g$ will be nearly 69 light-seconds. For 7.3 R$_g$ coronal size/height, the light travel time will be nearly 504 seconds, considering simple geometric effects. This matches with the measured reverberation time delay (483 $\pm$ 135 sec) between the direct powerlaw and soft excess during high flux, with 90\% confidence. The best-fit parameters of the lag spectral modelling during high flux provide a lower limit of the coronal size/height of 7.3 R$_g$. Therefore, our analysis of high flux variability indicates that the soft excess is possibly a reprocessed emission of the accretion disc with a separation of 7 R$_g$ or higher. During low flux, the origin is uncertain since the soft photons are delayed from hard photons by 96 seconds or less. Such a timescale is more consistent with the Comptonization timescale, where photons lose energy due to Compton downscattering.
Therefore, the analysis of mean spectra, Fourier-frequency dependent covariance spectra and time lag spectra implies that during the low flux state, either the corona is closer to the central supermassive black hole or has a very compact size (3.7 R$_g$ or less), which causes low direct, continuum flux. The disc must be truncated at a distance $>$1 R$_{ISCO}$ since (1) the Iron line is narrow, symmetric (top right panel of Figure \ref{fig2}), (2) the reflection continuum is present in covariance spectra (see Figure \ref{fig5}) of low flux at different timescales (from minutes to hours; the bottom left panel of Figure \ref{fig7}), (2) the disc ionisation is low ($log \xi$ = 0.36$^{+0.24}_{-0.11}$, see Table \ref{tab1}) (3) reverberation time delay is measured around 96 sec or less (see right panel of Figure \ref{fig10}). Variation of the reflection continuum at different time scales implies that a fraction of the reflection must come from the disc (the rest may come from the NLR region or a distant, cold reflector).
However, during high X-ray flux, the coronal height or spatial extension increases at least by a factor of $\sim$2, as suggested by our lag spectral analysis. Therefore, the high X-ray flux is expected from a corona which either moves away from the central black hole or has a larger spatial extension than that observed during low flux. In either case, a substantially larger photon flux is available as a direct continuum. At the same time, because of the increased number of direct photons, more reprocessed photons are expected to be observed from the disc. Hence, the X-ray flux due to the reprocessed continuum will also increase. Therefore, soft excess, which is considered a result of reflection in our work, is expected to be more significant during high flux. The enhanced reprocessing during high flux is expected to cause higher disc ionisation (which is observed in both mean (Table \ref{tab1}) and time-lag spectral analysis) and significantly higher variability in the reflected emission. Therefore, covariance spectra during high flux are predicted to be dominated by the variability of the reprocessed emission, which is also consistent with the results from covariance spectral analysis shown in the right panel of Figure \ref{fig7}.   
Our analysis, however, cannot distinguish whether the volumetric size of the corona grows during the low-to-high flux transition or if it is moved away along the vertical axis of the black hole. However, based on observed lag spectra fitted model-prediction, our estimation of coronal height during low and high fluxes indicates that disc-corona geometric configuration changes significantly, from smaller to larger extension, during the low-to-high flux transition.  

\section{Conclusions}\label{sec5}
By separating out the long (nearly 65ksec) high-flux X-ray observation of NGC 4593 from that of the low-flux observation, we have obtained a quantitative change in the disc corona coupled geometry system when the X-ray flux is changed by a factor of $\sim$2.5. We have used mean spectral analysis, Fourier-frequency resolved covariance spectral analysis, and energy-dependent lag spectral analysis of both high- and low-flux observations from \xmm{}. Our mean spectral analysis shows that (1) the absorption due to an ionised or neutral absorber does not change significantly from high to low flux, although both observations were taken apart by 14 years. (2) The narrow and symmetric Iron line complex implies the spectra are not complicated by Relativistic reflection. Hence, the change in flux is solely due to the change in the underlying continuum. During the high flux, an additional soft excess is observed, which can be modelled using a reflection model. Therefore, the question: `what is the origin of large X-ray flux variations in addition to the appearance of the soft X-ray excess during increased flux?' is the primary aim of the work. We address the question by employing two relatively unexplored techniques: Fourier frequency-resolved covariance spectral analysis and energy-dependent lag spectral modelling during both high- and low-flux. 
Covariance spectra are calculated in 9 different Fourier frequencies covering a time scale from 128 sec to 18 hours (equivalently, covering frequency ranges between 1.5-390 $\times$ 10$^{-5}$ Hz). Modelling of covariance spectra in all frequency ranges reveals that the direct powerlaw continuum dominates variabilities in all time scales during the low flux, while the same during the high flux is dominated by the reflection continuum most of the time. This may indicate that if reflection is caused by disc irradiations, then the amount of reflected emission compared to direct emission changes significantly during high to low flux transition.
As suggested by the lag spectral modelling, the size of the corona possibly shrinks significantly during low flux, or the corona moves closer to the black hole, which causes a significant decrease in the disc irradiation by coronal photons. During high flux, the corona possibly expands in size, causing an increase in direct powerlaw emission but reducing the powerlaw variabilities (1-5 keV fractional rms is significantly lower at low Fourier frequencies during high flux compared to that of the low flux as observed from Figure \ref{fig8}) while, at the same time, the irradiated flux variabilities and contribution increase. Such a hypothesis is supported by the appearance of the soft excess during the high flux, increased covariance normalisation of the reflection continuum over the powerlaw continuum and the detection of reverberation lag at the timescale of a couple of hundred seconds during high flux. Such a hypothesis may well be tested further using data from Seyfert galaxies showing relativistic reflections. However, such work is presently out of scope and can be deferred to future research. 

\section{Acknowldgements}
This paper is based on observations obtained with \xmm{}, an ESA science mission with instruments and contributions directly funded by ESA Member States and the USA (NASA). The research work at the Indian Institute of Technology, Hyderabad, is financially supported or funded by the Council of Scientific and Industrial Research (CSIR-Grant No: 09/1001(12694)/2021-EMR-I), Ministry of Science and Technology, Government of India. 
\section*{Data Availability}
The high-level data underlying this article are extracted through standard processing from raw data stored in HEASARC public archives.

\appendix
\section{Additional figures and tables}
\label{sec:append}
Covariance spectra extracted in different Fourier frequencies and fitted with the model combination of powerlaw and reflection during high and low fluxes are shown in Figure \ref{fig5a} and Figure \ref{fig5b}, respectively. Corresponding best-fit paramters are provided in Table \cref{taba1,taba2,taba3,taba4,taba5,taba6} respectively. 
Covariance spectra
\begin{figure*}
    \centering
    \includegraphics[scale=0.21,angle=-90]{2002-128-1seg-covfit.ps}
    \includegraphics[scale=0.21,angle=-90]{2002-128-3seg-covfit.ps}
    \includegraphics[scale=0.21,angle=-90]{2002-128-5seg-covfit.ps} \\
    \includegraphics[scale=0.21,angle=-90]{2002-256-1seg-covfit.ps}
    \includegraphics[scale=0.21,angle=-90]{2002-256-3seg-covfit.ps}
    \includegraphics[scale=0.21,angle=-90]{2002-256-5seg-covfit.ps}\\
    \includegraphics[scale=0.21,angle=-90]{2002-512-1seg-covfit.ps}
    \includegraphics[scale=0.21,angle=-90]{2002-512-3seg-covfit.ps}
    \includegraphics[scale=0.21,angle=-90]{2002-512-5seg-covfit.ps} \\
    \includegraphics[scale=0.21,angle=-90]{2002-1024-1seg-covfit.ps}
    \includegraphics[scale=0.21,angle=-90]{2002-1024-3seg-covfit.ps}
    \includegraphics[scale=0.21,angle=-90]{2002-1024-5seg-covfit.ps}
    \caption{Modelling of covariance spectra with during high flux: best-fit covariance energy spectra fitted with a combination of powerlaw and reflection continuum are shown along with model components and residuals for Fourier frequency segments S1B1, S4B1 and S6B1 (top left to top right panels), S1B2, S4B2 and S6B2 (upper middle left to upper middle right panels), S1B3, S4B3 and S6B3 (lower middle left to lower middle right panels) and S1B4, S4B4 and S6B4 (bottom left to bottom right panels) respectively. }
    \label{fig5a}
\end{figure*}

\begin{figure*}
\includegraphics[scale=0.21,angle=-90]{cov-fit-2016-1seg-128sec.ps}
\includegraphics[scale=0.21,angle=-90]{cov-fit-2016-3seg-128sec.ps}
\includegraphics[scale=0.21,angle=-90]{cov-fit-2016-5seg-128sec.ps} \\
\includegraphics[scale=0.21,angle=-90]{2016-256-1seg-covfit.ps}
\includegraphics[scale=0.21,angle=-90]{2016-256-3seg-covfit.ps}
\includegraphics[scale=0.21,angle=-90]{2016-256-5seg-covfit.ps} \\
\includegraphics[scale=0.21,angle=-90]{2016-512-1seg-covfit.ps}
\includegraphics[scale=0.21,angle=-90]{2016-512-3seg-covfitv2.ps}
\includegraphics[scale=0.21,angle=-90]{cov-fit-2016-5seg-512sec.ps} \\
\includegraphics[scale=0.21,angle=-90]{2016-1024-1seg-covfit.ps}
\includegraphics[scale=0.21,angle=-90]{2016-1024-3seg-covfit.ps}
\includegraphics[scale=0.21,angle=-90]{2016-1024-5seg-covfit.ps}
\caption{Modelling of covariance spectra with during low flux: best-fit covariance energy spectra fitted with a combination of powerlaw and reflection continuum are shown along with model components and residuals for the Fourier frequency segments S1B1, S4B1 and S6B1 (top left to top right panels), S1B2, S4B2 and S6B2 (upper middle left to upper middle right panels), S1B3, S4B3 and S6B3 (lower middle left to lower middle right panels) and S1B4, S4B4 and S6B4 (bottom left to bottom right panels) respectively. }
\label{fig5b}
\end{figure*}

\begin{table*}
\centering
\renewcommand{\arraystretch}{1.5}
\begin{tabular}{ccccc}
\hline
\hline
\thead{{\bf Model}} & & Fourier & frequency & segments \\ 
{\bf components} & 
\thead{{\bf Parameters}} & S1B2 & S4B2 & S6B2 \\
\hline
{\tt zpcfabs} & \thead{$N_H$ } & $0.82^{+0.21}_{-0.22}$ & $<0.16$ & -- \\

{\tt zpcfabs} & $F_{cvr}$ & $0.73^{+0.11}_{-0.05}$ & $<0.87$ & -- \\

{\tt xillver} & $\Gamma_{xill}$ & =$\Gamma_{PL}$ & =$\Gamma_{PL}$  & =$\Gamma_{PL}$ \\
{\tt xillver} & $A_{fe}$ & $4.5^{+0.54}_{-0.10}$ & $4.47^{+0.12}_{-0.9}$ & $<4.3$ \\

{\tt xillver} & \thead{$\log_{\xi}$ } & $3.19^{+0.54}_{-0.10}$ & $3.00^{+0.09}_{-0.07}$ & $2.91^{+0.28}_{-0.13}$ \\

{\tt xillver} & \thead{$norm$  ($\times10^{-2}$)} & $0.41^{+0.02}_{-0.03}$ & $0.52^{+0.01}_{-0.02}$  & $0.26^{+0.04}_{-0.04}$\\

{\tt powerlaw} & $\Gamma_{PL}$ & $2.4^{+0.11}_{-0.09}$ & $2.76^{+0.61}_{-0.29}$  &$2.38^{+0.38}_{-0.13}$ \\

{\tt powerlaw} & \thead{$norm$  ($\times10^{-2}$)} & $1.1^{+0.08}_{-0.03}$ & $2.14^{+0.64}_{-0.28}$ & $0.89^{+0.16}_{-0.04}$ \\

\hline
$\chi^2/dof$ &  & \thead{$23.71/16$  } & \thead{$18.90/16$ } & \thead{$20.61/18$ } \\
\hline
\end{tabular}
\caption{Bestfit parameters for 0.5-9.0 keV covariance photon energy spectra extracted for S1B2, S4B2 and S6B2 segments (see Table \ref{tab-freq}) during low flux are provided along with their models and 1$\sigma$ errors. $F_{cvr}$ is the covering fraction. The unit of N$_H$, Hydrogen column density is $10^{22}\ cm^{-2}$, $\xi$ is the ionization parameter with the unit of $ergs\ cm^{-2}\ s^{-1}$. The high energy cutoff, disc inclination angle and reflection fraction are kept fixed at 300 keV, 56$^o$ and -1.0, respectively, in all fittings. }
\label{taba1}
\end{table*}

\begin{table*}
\centering
\renewcommand{\arraystretch}{1.5}
\begin{tabular}{ccccc}
\hline
\thead{{\bf Model}} & & Fourier & frequency & segments \\ 
{\bf components} & 
\thead{{\bf Parameters}} & S1B3 & S4B3 & S6B3 \\
\hline
{\tt zpcfabs} & \thead{$N_H$  } & $0.76^{+0.19}_{-0.08}$ & -- & -- \\

{\tt zpcfabs} & $CvrFract$ & $0.72^{+0.04}_{-0.05}$ & -- & -- \\

{\tt xillver} & $\Gamma_{xill}$ & =$\Gamma_{PL}$ & =$\Gamma_{PL}$  & =$\Gamma_{PL}$ \\
{\tt xillver} & $A_{fe}$ & $4.5(f)$ & $4.5(f)$ & $4.5^{+0.21}_{-0.05}$ \\

{\tt xillver} & \thead{$\log_{\xi}$} & $3.07^{+0.09}_{-0.19}$ & $2.82^{+0.69}_{-0.60}$ & $2.81^{+0.31}_{-0.05}$ \\

{\tt xillver} & \thead{$norm$  ($\times10^{-2}$)} & $0.46^{+0.06}_{-0.09}$ & $0.17^{+0.02}_{-0.06}$  & $0.14^{+0.05}_{-0.08}$\\

{\tt powerlaw} & $\Gamma_{PL}$ & $2.41^{+0.13}_{-0.19}$ & $1.83^{+0.07}_{-0.15}$  &$1.8^{+0.20}_{-0.06}$ \\

{\tt powerlaw} & \thead{$norm$  ($\times10^{-2}$)} & $3.38 ^{+0.12}_{-0.93}$ & $1.05^{+0.42}_{-0.45}$ & $1.03 ^{+0.73}_{-0.38}$ \\

\hline
$\chi^2/dof$ &  & \thead{$28.57/16$  } & \thead{$15.99/18$ } & $\thead{20.68/18}$  \\
\hline
\end{tabular}
\caption{Bestfit parameters for 0.5-9.0 keV covariance photon energy spectra extracted for S1B3, S4B3 and S6B3 segments (see Table \ref{tab-freq}) during low flux are provided along with their models and 1$\sigma$ errors. $F_{cvr}$ is the covering fraction. The unit of N$_H$, Hydrogen column density is $10^{22}\ cm^{-2}$, $\xi$ is the ionization parameter with the unit of $ergs\ cm^{-2}\ s^{-1}$. The high energy cutoff, disc inclination angle and reflection fraction are kept fixed at 300 keV, 56$^o$ and -1.0, respectively, in all fittings. }
\label{taba2}
\end{table*}

\begin{table*}
\centering
\renewcommand{\arraystretch}{1.5}
\begin{tabular}{ccccc}
\hline
\thead{{\bf Model}} & & Fourier & frequency & segments \\ 
{\bf components} & 
\thead{{\bf Parameters}} & S1B4 & S4B4 & S6B4 \\
\hline
{\tt zpcfabs} & \thead{$N_H$} & $0.75^{+0.21}_{-0.13}$ & $<0.11$ & -- \\

{\tt zpcfabs} & $F_{cvr}$ & $0.75^{+0.12}_{-0.07}$ & $0.75(f)$ & -- \\

{\tt xillver} & $\Gamma_{xill}$ & =$\Gamma_{PL}$ & =$\Gamma_{PL}$  & =$\Gamma_{PL}$ \\
{\tt xillver} & $A_{fe}$ & 3.5(f) & $3.5^{+0.}_{-1.7}$ & $3.49^{+0.0}_{-1.2}$ \\

{\tt xillver} & \thead{$\log_{\xi}$  } & $2.92^{+0.0}_{-0.0}$ & $3.06^{+0.19}_{-0.11}$ & $2.99^{+0.26}_{-0.10}$ \\

{\tt xillver} & \thead{$norm$  ($\times10^{-2}$)} & $0.47^{+0.06}_{-0.04}$ & $0.39^{+0.08}_{-0.07}$  & $0.06^{+0.03}_{-0.01}$\\

{\tt powerlaw} & $\Gamma_{PL}$ & $2.29^{+0.18}_{-0.11}$ & $2.53^{+0.43}_{-0.23}$  &$2.01^{+0.2}_{-0.08}$ \\

{\tt powerlaw} & \thead{$norm$  ($\times10^{-2}$)} & $3.2 ^{+0.19}_{-0.65}$ & $1.62^{+0.31}_{-0.85}$ & $1.18^{+0.72}_{-0.42}$ \\

\hline
$\chi^2/dof$ & & 27.30/16 & 18.16/16 & 16.89/18 \\
\hline

\hline
\end{tabular}
\caption{Bestfit parameters for 0.5-9.0 keV covariance photon energy spectra extracted for S1B4, S4B4 and S6B4 segments (see Table \ref{tab-freq}) during low flux are provided along with their models and 1$\sigma$ errors. $F_{cvr}$ is the covering fraction. The unit of N$_H$, Hydrogen column density is $10^{22}\ cm^{-2}$, $\xi$ is the ionization parameter with the unit of $ergs\ cm^{-2}\ s^{-1}$. The high energy cutoff, disc inclination angle and reflection fraction are kept fixed at 300 keV, 56$^o$ and -1.0, respectively, in all fittings. }
\label{taba3}
\end{table*}

\begin{table*}
\centering
\renewcommand{\arraystretch}{1.5}
\begin{tabular}{ccccc}
\hline
\thead{{\bf Model}} & & Fourier & frequency & segments \\ 
{\bf components} & 
\thead{{\bf Parameters}} & S1B2 & S4B2 & S6B2 \\
\hline
{\tt zpcfabs} & \thead{$N_H$}   & $0.53^{+0.29}_{-0.08}$ & -- & -- \\

{\tt zpcfabs} & $F_{cvr}$ & $0.75^{+0.08}_{-0.05}$ & -- & -- \\

{\tt xillver} & $\Gamma_{xill}$ & =$\Gamma_{PL}$ & =$\Gamma_{PL}$  & =$\Gamma_{PL}$ \\
{\tt xillver} & $A_{fe}$ & $2.62^{+0.71}_{-0.85}$ & $4.5^{+0.14}_{-0.92}$ & $<4.5$ \\

{\tt xillver} & \thead{$\log_{\xi}$} & $4.04^{+0.52}_{-0.70}$ & $3.52^{+0.24}_{-0.14}$ & $3.7^{+0.28}_{-0.06}$ \\

{\tt xillver} & \thead{$norm$  ($\times10^{-2}$)} & $0.89^{+0.05}_{-0.03}$ & $0.16^{+0.02}_{-0.02}$  & $0.11^{+0.02}_{-0.03}$\\

{\tt powerlaw} & $\Gamma$ & $2.99 ^{+0.30}_{-0.36}$ & $2.35^{+0.059}_{-0.09}$  &$2.54^{+0.08}_{-0.04}$ \\

{\tt powerlaw} & \thead{$norm$  ($\times10^{-2}$)} & $0.34^{+0.08}_{-0.02}$ & $0.14^{+0.05}_{-0.02}$ & $0.07^{+0.03}_{-0.01}$ \\

\hline
$\chi^2/dof$ &  & \thead{$15.66/16$} & \thead{$17.91/19$} &\thead{ $20.47/18$} \\
\hline
\end{tabular}

\caption{Bestfit parameters for 0.5-9.0 keV covariance photon energy spectra extracted for S1B2, S4B2 and S6B2 segments (see Table \ref{tab-freq}) during high flux are provided along with their models and 1$\sigma$ errors. $F_{cvr}$ is the covering fraction. The unit of N$_H$, Hydrogen column density is $10^{22}\ cm^{-2}$, $\xi$ is the ionization parameter with the unit of $ergs\ cm^{-2}\ s^{-1}$. The high energy cutoff, disc inclination angle and reflection fraction are kept fixed at 300 keV, 56$^o$ and -1.0, respectively, in all fittings.}
\label{taba4}
\end{table*}

\begin{table*}
\centering
\renewcommand{\arraystretch}{1.5}
\begin{tabular}{ccccc}
\hline
\thead{{\bf Model}} & & Fourier & frequency & segments \\ 
{\bf components} & 
\thead{{\bf Parameters}} & S1B3 & S4B3 & S6B3 \\
\hline
{\tt zpcfabs} & \thead{$N_H$ } & $0.37^{+0.11}_{-0.07}$ & -- & -- \\

{\tt zpcfabs} & $F_{cvr}$ & $0.65^{+0.05}_{-0.06}$ & -- & -- \\

{\tt xillver} & $\Gamma_{xill}$ & =$\Gamma_{PL}$ & =$\Gamma_{PL}$  & =$\Gamma_{PL}$ \\
{\tt xillver} & $A_{fe}$ & $2.91^{+0.43}_{-0.67}$ & $<3.89$ & $4.33^{+0.12}_{-0.07}$ \\

{\tt xillver} & \thead{$\log_{\xi}$ } & $3.61^{+0.11}_{-0.24}$ & $3.32^{+0.25}_{-0.15}$ & $3.38^{+0.16}_{-0.21}$ \\

{\tt xillver} & \thead{$norm$  ($\times10^{-2}$)} & $0.89^{+0.05}_{-0.03}$ & $0.17^{+0.03}_{-0.02}$  & $0.13^{+0.03}_{-0.04}$\\

{\tt powerlaw} & $\Gamma_{PL}$ & $2.82 ^{+0.19}_{-0.16}$ & $2.26^{+0.13}_{-0.11}$  &$2.21^{+0.08}_{-0.09}$ \\

{\tt powerlaw} & \thead{$norm$  ($\times10^{-2}$)} & $0.28^{+0.056}_{-0.029}$ & $0.94^{+0.01}_{-0.01}$ & $0.09^{+0.02}_{-0.39}$  \\

\hline
$\chi^2/dof$ & & 15.18/16 & 15.67/18 & 12.36/18 \\
\hline
\end{tabular}
\caption{Bestfit parameters for 0.5-9.0 keV covariance photon energy spectra extracted for S1B3, S4B3 and S6B3 segments (see Table \ref{tab-freq}) during high flux are provided along with their models and 1$\sigma$ errors. $F_{cvr}$ is the covering fraction. The unit of N$_H$, Hydrogen column density is $10^{22}\ cm^{-2}$, $\xi$ is the ionization parameter with the unit of $ergs\ cm^{-2}\ s^{-1}$. The high energy cutoff, disc inclination angle and reflection fraction are kept fixed at 300 keV, 56$^o$ and -1.0, respectively, in all fittings.}
\label{taba5}
\end{table*}

\begin{table*}
\centering
\renewcommand{\arraystretch}{1.0}
\begin{tabular}{ccccc}
\hline
\thead{{\bf Model}} & & Fourier & frequency & segments \\ 
{\bf components} & 
\thead{{\bf Parameters}} & S1B4 & S4B4 & S6B4 \\
\hline
{\tt zpcfabs} & \thead{$N_H$} & $0.51^{+0.10}_{-0.04}$ & $-$ & $-$ \\

{\tt zpcfabs} & $F_{cvr}$ & $0.77^{+0.06}_{-0.10}$ & $-$ & $-$ \\

{\tt xillver} & $\Gamma_{xill}$ & =$\Gamma_{PL}$ & =$\Gamma_{PL}$  & =$\Gamma_{PL}$ \\

{\tt xillver} & $A_{fe}$ & $4.5(f)$ & $4.5(f)$ & $<$4.5 \\

{\tt xillver} & \thead{$\log_{\xi}$  } & $3.39^{+0.05}_{-0.31}$ & $3.67^{+0.12}_{-0.10}$ & $2.87^{+0.18}_{-0.57}$ \\

{\tt xillver} & \thead{$norm$  ($\times10^{-2}$)} & $1.28^{+0.52}_{-0.32}$ & $0.13^{+0.03}_{-0.02}$  & $0.16^{+0.07}_{-0.03}$\\

{\tt powerlaw} & $\Gamma_{PL}$ & $2.89^{+0.22}_{-0.34}$ & $2.39^{+0.05}_{-0.11}$  &$2.19^{+0.08}_{-0.18}$ \\

{\tt powerlaw} & \thead{$norm$  ($\times10^{-2}$)} & $0.12^{+0.06}_{-0.03}$ & $0.04^{+0.01}_{-0.01}$ & $<$0.09  \\

\hline
$\chi^2/dof$ & & 14.21/16 & 16.50/18 & 8.81/18 \\
\hline

\hline
\end{tabular}
\caption{Best-fit parameters for 0.5-9.0 keV covariance spectra extracted for S1B4, S4B4 and S6B4 segments (see Table \ref{tab-freq}) during high flux are provided along with their models and 1$\sigma$ errors. $F_{cvr}$ is the covering fraction. The unit of N$_H$, Hydrogen column density is $10^{22}\ cm^{-2}$, $\xi$ is the ionization parameter with the unit of $ergs\ cm^{-2}\ s^{-1}$. The high energy cutoff, disc inclination angle and reflection fraction are kept fixed at 300 keV, 56$^o$ and -1.0, respectively, in all fittings.}
\label{taba6}
\end{table*}

\end{document}